# HIERARCHICAL PARTITION-BASED ANONYMOUS ROUTING PROTOCOL (HPAR) IN MANET FOR EFFICIENT AND SECURE TRANSMISSION

Fahmida Aseez[1] and Dr.Sheena mathew[2]

[1]Mtech Student Division of Computer Engineering, SOE, CUSAT, Cochin, India
Fahmida044@gmail.com
[2]Dr.Sheena Mathew, Professor, Division of Computer Engineering, SOE, CUSAT, Cochin, India
sheenamathew@cusat.ac.in

## ABSTRACT

*Anonymous routing protocols are used in MANET's to hide the nodes from outsiders in order to protect from various attacks. HPAR partitions the network area dynamically into zones and chooses nodes in zones randomly as intermediate relay nodes .This relay nodes help in secure routing. In HPAR anonymity protection is given to source, destination and route. HPAR have low cost and provide high level of protection. It has techniques to counter various attacks.*

## 1. INTRODUCTION

MANET Comprises of wireless mobile nodes that are freely and self-organize into a temporary network topology with-out any infrastructural support. Open nature, dynamic changing topology, no central management are the key features of MANET's. They have a variety of applications in military, banking, education, commerce etc. Security in MANET is a major issue. Data get lost or stolen by tampering and analyzing data and traffic analysis by eavesdropping method or attacking routing protocol. Solution to this is anonymous routing.

In MANET, the term Anonymity means hiding identity of the source node, receiver node and the chosen path. Anonymous routing protocols provide secure communications by hiding node identities and preventing traffic analysis attacks from outside observers. They are used in Military, Banking like application, where security of communication is a major concern. Anonymity is critical in military applications for example soldier communication. MANET deployed in a battlefield can be vulnerable to traffic analysis; enemies may intercept transmitted packets, track our soldiers (i.e., nodes), attack the commander nodes, and block the data transmission etc.

Limited resource is an inherent problem in MANETs. MANETs' complex routing and strict channel resource constraints impose strict limits on the system capacity. Nowadays multimedia applications (e.g., video transmission) require high routing efficiency. Our existing anonymous routing protocols [1] generate a significantly high cost, which badly affect the resource constraint problem in MANETs. A MANET employed in a battlefield, with a high-cost anonymous routing and low quality of service in voice and video data transmission due to depleted resources may lead to disastrous delay in military operations.

HPAR provide high anonymity protection (for sources, destination, and route) with low cost. HPAR dynamically partitions a network field into zones and randomly chooses nodes in zones as intermediate relay nodes, which form a non-traceable anonymous route. In each routing step, a data sender or forwarder partitions the network field in order to separate itself and the destination into two zones. It then randomly chooses a node in the other zone as the next relay node and use GPSR algorithm to send the data to the relay node. At last, the data is broadcasted to k nodes in the destination zone, providing k-anonymity to the destination. Also it has strategy to hide the data initiator among a number of initiators to strengthen the anonymity protection of the source. HPAR is also resilient to intersection attacks and timing attacks.

## 2. LITERATURE SURVEY

Anonymity of a subject [2] means that the subject is not identifiable within a set of subjects, the anonymity set. Anonymity of a subject from an attacker's perspective means that the attacker cannot sufficiently identify the subject within a set of subjects, the anonymity set. In simple terms Anonymity provides the privacy protection in the communication.

An existing protocol ALARM [3] is a table driven protocol, with location based routing. ALARM provides Security against active and passive attacks by advanced cryptographic techniques such as group signature. Group signature ensures that only valid members who have registered with the group manager can decrypt and read the packets. This protocol sends out Location Announcement Messages (LAM) to inform all the nodes of the network topology from time to time. Problem with ALARM is cannot protect location anonymity of source and destination node.

To preserve privacy PRISM [4] protocol suggested the use of Location bases routing along with Group signatures. It is an on-demand routing scheme. A source node will initiate a route discovery phase when it has data to transmit. PRISM is Based on the concept of location aided routing it located the destination, encrypts he packet, insert the source group signature and send the packet. Receiving packets can verify the group signature and destination is identified with the coordinates. The Route reply consists of a session key which will be used for further communication for that particular session. These Routes are discarded after communication. This protocol achieves privacy and security against active as well as passive attacks. As the nodes identity is not revealed and the destination node location is encrypted by key known only to valid group members.ALARM is a link-state protocol and exposes the entire topology to all insiders While PRISM prevents inside attacks.

Many anonymity routing algorithms are based on the geographic routing protocol for e.g., Greedy Perimeter Stateless Routing (GPSR) [5]. GPSR, packets are routed geographically. GPSR can route a packet to any connected destination. There are two distinct algorithms GPSR uses for routing first a greedy forwarding algorithm that moves packets progressively closer to the destination at each hop, and a perimeter forwarding algorithm that forwards packets where greedy forwarding is impossible. The greedy forwarding rule is simple: a node x forwards a packet to its neighbor y that is closest to the destination D as shown in Figure 1. Greedy forwarding fails when no neighbor is closer than *x* to the destination. GPSR recovers from greedy forwarding failure using perimeter mode, which amounts to forwarding packets using the right-hand rule shown in Figure 2.

AO2P *(*ad hoc on-demand position-based private routing Algorithm) Protocol [6] is mainly proposed for communication anonymity. Route discovery is done by using only the position of the destination. Other information such as forwarding nodes positions are hiding from the network. [7] Provides an insight about Traffic Analysis. If the different routes that can be taken require different amounts of time, the system could be vulnerable to timing attacks. Intersection attacks mainly occurs by An attacker having information about what users are active at any given time can, through repeated observations, determine what users communicate with each other.

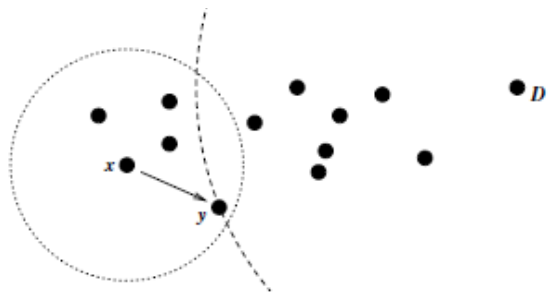

Figure 1. Greedy forwarding example x forwards to y which is closest to D.

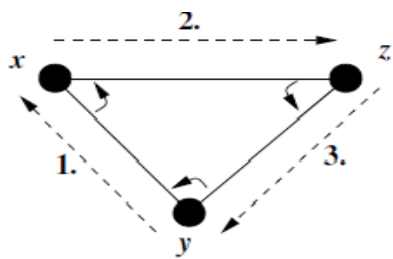

Figure 2. Right hand rule example packets travel along clockwise around the enclosed region.

Various security attacks include passive and active attacks [8]. A passive attack does not alter the data transmitted within the network. Active attacks are very severe attacks on the network that prevent message flow between the nodes. Active attacks are classified into three groups: 1) Dropping Attacks Compromised nodes or selfish nodes can drop all packets that are not destined for them. Dropping attacks can prevent end-to-end communications between nodes. 2) Modification Attacks modify packets and disrupt the overall communication between network nodes. Sinkhole attacks are the example of modification attacks. 3) Fabrication Attacks the attacker send fake message to the neighboring nodes without receiving any related message.

## 3. PROPOSED SYSTEM

HPAR partitions given network area into two zones as horizontally (or vertically). Then again split every partition into two zones as vertically (or horizontally). This process called as hierarchical zone partition Figure 3 [1]. After partitioning HPAR randomly select a node in each zone at each step as an intermediate relay node. While this partitioning each data source of forwarder node checks whether itself and destination nodes are not in same zone. If it is not then partitioning continues. While in routing first source node randomly chooses a node in other zone known as temporary destination (TD) .Then uses GPSR routing algorithm to send the data to node close to TD. A node closer to TD known as Random Forwarder (RF). This repeats until destination zone is reached. But in destination zone data is broadcasted in ZD to k nodes which makes attacker or observer does not know the destination node. For successful completion of data transmission destination node send a confirmation to source node. If source node not receives to confirm during predefined time period, it will resend packets.

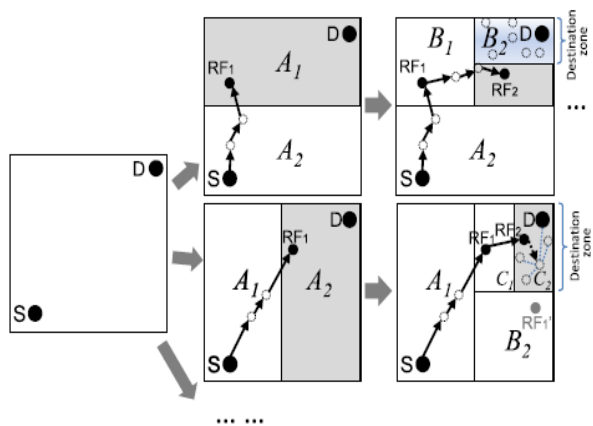

Figure 3. Zone partitioning

Different modules of the system includes Node construction, Zone partition, Source anonymity, Routing protocol , Destination anonymity Figure 4 shows the proposed system architecture.

Routing steps:
Step1: Assume rectangle network area, nodes are disseminated.
Step2: Each data source or forwarder executes the hierarchical zone partition
Step3: First check whether itself and D are in same zone.
Step4: If so, then divides the zone partition as Hierarchical zone partition.
Step5: Repeat step 4 process until itself and ZD are not in zone.
Step6: If source and ZD are not in the same zone then it randomly chooses a position in the other zone is called TD (Temporary Destination).
Step7: Using GPSR to send the data to the node closest to TD. This node is defined as a RF (Random Forwarder).
Step8: Repeat step 6 and step 7 until a data receiver finds itself residing in ZD having k node
Step9: In the last step, the data is broadcasted to k nodes in the destination zone, providing k-anonymity to the D.

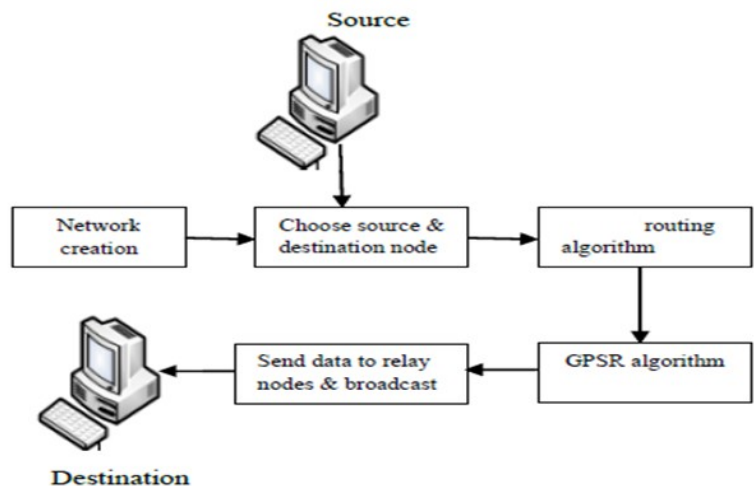

Figure 4. Proposed system architecture

A source node S sends a request to a destination node D and the destination responds with data. Each node uses a dynamic pseudonym as its node identifier rather than using its real MAC address, which can be used to trace nodes existence in the network. To avoid pseudonym collision, we use a collision Resistant hash function, such as SHA-1, to hash a node's MAC address and current time stamp. Each node periodically piggybacks its updated position and pseudonym to "hello" messages, and sends the messages to its neighbors. Also, every node maintains a routing table that keeps its neighbors pseudonyms associated with their locations.

Destination zone position is calculated by using certain equations. Zone position refers to the upper left and bottom-right coordinates of a zone.

$$H = log_2\left(\frac{\rho \cdot G}{k}\right),$$

- Let H denote the total number of partitions in order to produce ZD. Using the number of nodes in ZD (i.e., k), and node density.
- k = number of nodes in ZD
- P = node density
- G = the size of the entire network area.
- Using the calculated H, the size G, the positions (0,0) and $(X_g, Y_g)$ of the entire network area, and the position of D, the source S can calculate the zone position of ZD

Therefore, the size of the destination zone is given as:

$$\frac{G}{2^H}.$$

## 3.1 Anonymity Protection

HPAR makes the route between a S-D pair difficult to discover by randomly and dynamically selecting the relay nodes. The resultant different routes for transmissions between a given S-D pair make it difficult for an intruder to observe a statistical pattern of transmission. HPAR incorporates the "notify and go" mechanism to prevent an intruder from identifying which node within the source neighbourhood has initiated packets. HPAR also provides k-anonymity to destinations by hiding D among k receivers in Z d. Thus, an eavesdropper can only obtain information on Z d, rather than the destination position, from the packets and nodes en route.

In HPAR the nodes entire network is grouped to form clusters as in Figure 5. The clustering is based on the position or the coordinates of the nodes. Distance between the nodes and the source is calculated. Based on the distance the nodes are grouped. The nodes that are in the specified distance are forming a cluster. Then the communication is in the name of these clusters or groups. The packet transmission is by the communication between the groups. Inter and intra group communication is by random forwarders and relay nodes. So the communication is multi hop clustering. Each cluster has a specified range. The nodes belonging to that range are determined by the basics of their distance or coordinates. Each cluster maintains a cluster identifier or group identifier. The communication is carried out in this group id. The nodes form a cluster if they belong to particular range or distance. The nodes within the group can communicate with each other. This is known as intra group routing .They are mostly neighbors or one hop nodes. The nodes outside the group are communicated as multi hop fashion. The routing between the groups are known as inter group routing. This is by the means of relay nodes and random forwarders.

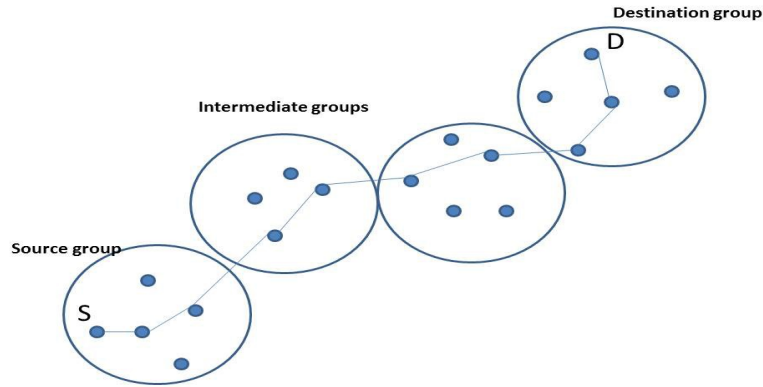

Figure 5. Clustering in HPAR

The HPAR uses the hierarchical clustering scheme and randomly chooses a node in the cluster or group in each step as an intermediate relay node as random forwarder. The source group consists of the sender. It transmits the packets to next random forwarder from that group or next group. The random forwarder in the next group can understand that the packet is from that group that node can't get the idea of real sender. After passing through the intermediate nodes it finally reaches the destination cluster node. Then it forwards to exact destination. In the clustering the communication between the clusters are by the name of cluster group identifier. So the real identity of each node inside the cluster is maintained. The communications between the clusters are by the group or cluster identifier. Each group maintained and identifier. So the outside communication hides the real identity of the node by this group identity.

### 3.2 Strategies against attacks

3.2.1 Resilience to Timing Attacks

Two nodes A and B communicate with each other at an interval of 5 seconds. After a long observation time, the intruder finds that A's packet sending time and B's packet receiving time have a fixed five second difference. Then, the intruder would suspect that A and B are communicating with each other. Avoiding the exhibition of interaction between communication nodes is a way to counter timing attacks.

3.2.2 Strategy to Counter Intersection Attacks

In intersection attack an attacker with information about active users at a given time can determine the sources and destinations that communicate with each other through repeated observations. rather than using direct local broadcasting in the zone, the last RF multicasts packet pkt 1 to a partial set of nodes m.The m nodes hold the packets until the arrival of the next packet pkt 2 .Upon the arrival of the next packet, the m nodes conduct one-hop broadcasting to enable other nodes in the zone to also receive the packet in order to hide D.

Comparison of HPAR, ALARM and AO2P protocols based on some parameters:

1) Number of actual participating nodes

ALARM and AO2P is based on the GPSR method. GPSR always proceeds through the shortest paths. So the number of actual participating nodes is less compared to HPAR.

2) Latency in packet transmission

Latency is defined as the time difference between the packet transmissions and receiving. Latency in HPAR is significantly lower than the other two. This is because of the time needed for the public key encryption of ALARM and AO2P. HPAR follows symmetric key encryption only once which reduces the latency.

3) Packet delivery rate

Fraction of successfully delivered packets to a destination is called the delivery rate .HPAR has higher delivery rates compared to AO2P and ALARM, as a result of final local broadcast process HPAR achieves enhanced route anonymity than ALARM and AO2P. HPAR has more number of actual participating nodes and its random relay node selection boost the anonymity.

4) Transmission cost

Transmission cost and latency in packet transmission are lower in HPAR compared with the other two. HPAR contributes better data delivery rate than ALARM and AO2P.

## 4. CONCLUSIONS

Anonymous routing protocols are crucial in MANETs to provide secure communications by hiding node identities and routes from outside observers. Anonymity in MANETs includes identity and location anonymity of senders and destinations as well as route anonymity. The aim is to make the communication between different nodes anonymous in MANET. By anonymity we mean that intermediate nodes are unaware of the sender and destination. Only the sender will know the receiver and only the receiver will know the sender. HPAR can offer high anonymity protection at a low cost when compared to other anonymity algorithms.It can also achieve comparable routing efficiency also Provide Resilience to Timing Attacks and Strategy to Counter Intersection Attacks.

### ACKNOWLEDGEMENTS

The authors would like to thank everyone.

Fahmida Aseez

Post-graduate student at school of Engineering CUSAT. Had graduated from Adi Sankara Institute of Engineering and Technology Kalady. Area of interest is in Network Computing.

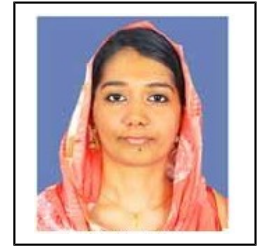

Dr.Sheena Mathew

Professor in Division of Computer Science, SOE, Cochin University has 23 years of teaching experience in Computer Science. She had her graduation from Madurai Kamaraj University, post-graduation from Indian Institute of Science,Banglore and doctorate from CUSAT. She was the head of Department of Division of Computer Science and Engineering, school of engineering, CUSAT for the period of four years. Her areas of interest being Cryptography and Network Security. She has more than 30 publications in various international journals and conferences to her credit.